\documentclass{svjour3}                     
\smartqed  
\usepackage{graphicx}
\journalname{Experimental Astronomy}
\begin{document}

\title{First results from a next-generation off-plane X-ray diffraction grating}

\titlerunning{Next-generation off-plane gratings}        

\author{Randall McEntaffer, Casey DeRoo, Ted Schultz, Brennan Gantner, James Tutt, Andrew Holland, Stephen O'Dell, Jessica Gaskin, Jeffrey Kolodziejczak, William W. Zhang, Kai-Wing Chan, Michael Biskach, Ryan McClelland, Dmitri Iazikov, Xinpeng Wang, \& Larry Koecher}


\institute{R. McEntaffer \at
              Department of Physics \& Astronomy, University of Iowa, Van Allen Hall, Iowa City, IA 52242 \\
              Tel.: +001-319-335-3007\\
              \email{randall-mcentaffer@uiowa.edu}           
           \and
           C. DeRoo \at
              Department of Physics \& Astronomy, University of Iowa, Van Allen Hall, Iowa City, IA 52242
              \and
           T. Schultz \at
              Department of Physics \& Astronomy, University of Iowa, Van Allen Hall, Iowa City, IA 52242
              \and
           B. Gantner \at
              Department of Physics \& Astronomy, University of Iowa, Van Allen Hall, Iowa City, IA 52242
              \and   
           J. Tutt \at
              Centre for Electronic Imaging, The Open University, Walton Hall, Milton Keynes, MK7 6AA, UK 
              \and
           A. Holland \at
              Centre for Electronic Imaging, The Open University, Walton Hall, Milton Keynes, MK7 6AA, UK
              \and
           S. O'Dell \at
              NASA Marshall Space Flight Center, 320 Sparkman Dr. NW, Huntsville, AL 35805
              \and
           J. Gaskin \at
              NASA Marshall Space Flight Center, 320 Sparkman Dr. NW, Huntsville, AL 35805
              \and
           J. Kolodziejczak \at
              NASA Marshall Space Flight Center, 320 Sparkman Dr. NW, Huntsville, AL 35805
              \and
           W. Zhang \at
              NASA Goddard Space Flight Center, Bldg 34, Greenbelt, MD 20771
              \and
           K. Chan \at
              NASA Goddard Space Flight Center, Bldg 34, Greenbelt, MD 20771
              \and
           M. Biskach \at
              NASA Goddard Space Flight Center, Bldg 34, Greenbelt, MD 20771
              \and
           R. McClelland \at
              NASA Goddard Space Flight Center, Bldg 34, Greenbelt, MD 20771
              \and                     
           D. Iazikov \at
              LightSmyth Technologies, Inc., 875 Wilson St. Suite C, Eugene, OR 97402
              \and
           X. Wang \at
              Nanonex Corporation, 1 Deerpark Dr., Monmouth Jct., NJ 08852
              \and
           L. Koecher \at
              Nanonex Corporation, 1 Deerpark Dr., Monmouth Jct., NJ 08852
}

\date{Received: date / Accepted: date}

\maketitle

\begin{abstract}
Future NASA X-ray spectroscopy missions will require high throughput, high resolution grating spectrometers.  Off-plane reflection gratings are capable of meeting the performance requirements needed to realize the scientific goals of these missions.  We have identified a novel grating fabrication method that utilizes common lithographic and microfabrication techniques to produce the high fidelity groove profile necessary to achieve this performance.  Application of this process has produced an initial pre-master that exhibits a radial (variable line spacing along the groove dimension), high density ($>$6000 grooves/mm), laminar profile.  This pre-master has been tested for diffraction efficiency at the BESSY II synchrotron light facility and diffracts up to 55\% of incident light into usable spectral orders.  Furthermore, tests of spectral resolving power show that these gratings are capable of obtaining resolutions well above 1300 ($\lambda/\Delta\lambda$) with limitations due to the test apparatus, not the gratings.  Obtaining these results has provided confidence that this fabrication process is capable of producing off-plane reflection gratings for the next generation of X-ray observatories.
\end{abstract}

\section{Introduction}
\label{intro}
Soft X-ray energies harbor a wealth of transition lines useful for performing key diagnostics in astrophysical plasmas.  Grating spectrometers are employed on current X-ray observatories, such as \textit{XMM-Newton} and the \textit{Chandra X-ray Observatory}, and had been baselined for the recently cancelled \textit{International X-ray Observatory} (\textit{IXO}) program to obtain spectra in this regime. Next-generation observatories, however, will require significant improvements in spectrometer performance in order to accomplish future science goals.  For example, \textit{IXO}'s X-ray Grating Spectrometer (XGS) sought spectral resolutions of $\Delta\lambda /\lambda > $ 3000 and effective areas of $\sim$1000 cm$^2$ for an energy range of 0.3--1.0 keV. This represents nearly a full order of magnitude increase in both specifications from the grating instruments of \textit{Chandra} and \textit{XMM-Newton}, thus necessitating the development of a new generation of high quality spectrometers capable of achieving these performance requirements.

Spectrometers employing off-plane reflection gratings are a promising approach to soft x-ray spectroscopy, offering preferable telescope packing geometries, excellent throughput, and the potential for high resolution in the desired bands. These spectrometers typically use a set of nested Wolter-I optics (a primary parabolic mirror, followed by a secondary hyperbolic), focusing x-rays over a length of several meters. A short distance down the optical axis, an array of gratings positioned in the off-plane mount disperses the converging light \cite{McEntaffer2010}. This dispersion is in the shape of a cone (giving the common name conical diffraction) and forms an arc of diffracted light at the focal plane. The dispersed spectrum is then imaged by a detector, typically a CCD camera, placed at the focal plane. 

\begin{figure}
  \includegraphics[width=5.0in,height=2.63in]{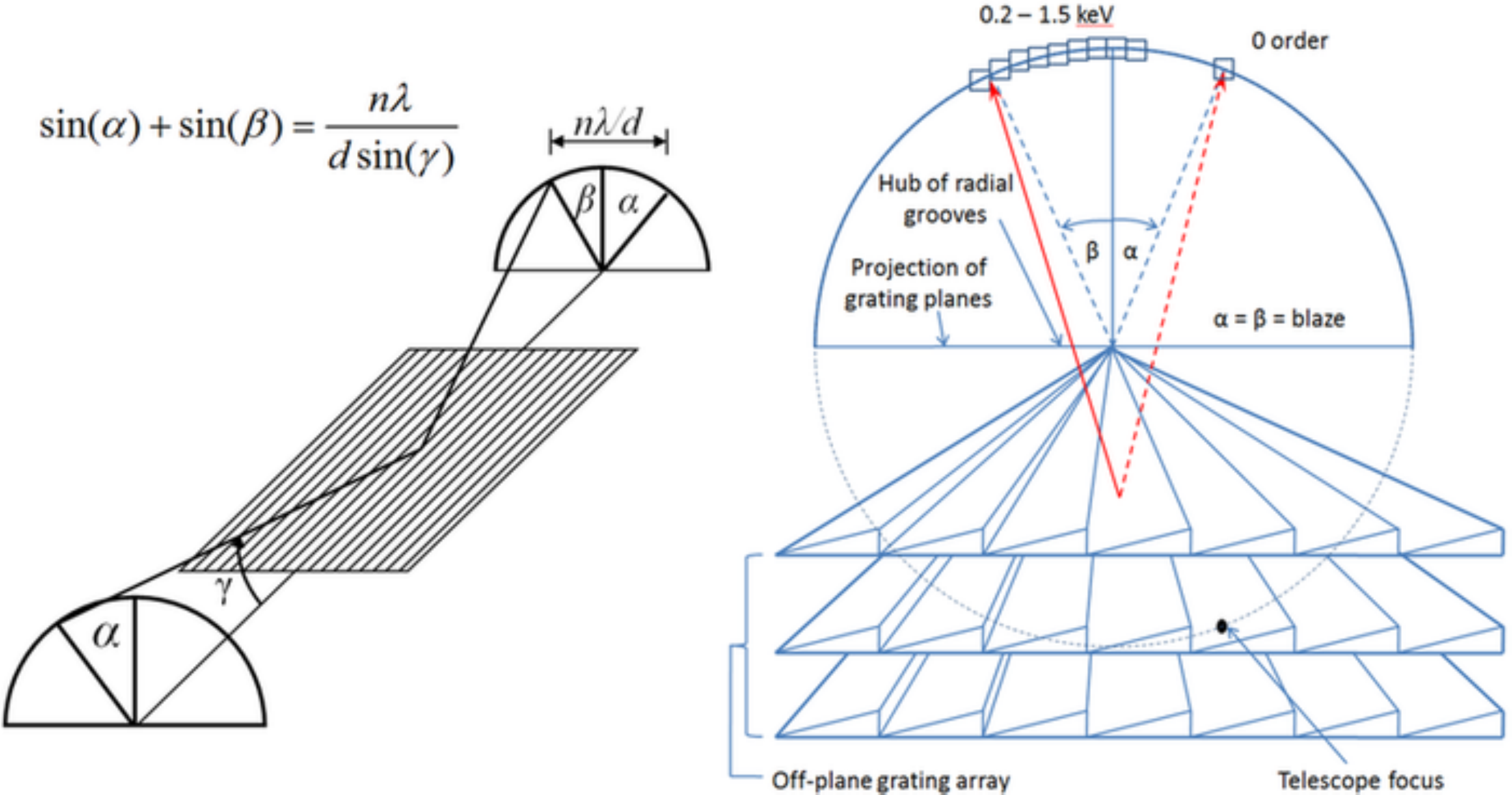}
\caption{\textit{Left} - The off-plane grating mount. \textit{Right} - Three gratings, placed many meters from the focus are shown projected onto the focal plane to elucidate the nature of the arc of diffraction which is detected by an array of CCDs depicted as squares.}
\label{OPGeom}
\end{figure}

The off-plane mount requires customized gratings in order to be optimized for high resolutions and increased efficiencies. Figure \ref{OPGeom} shows the grating geometry and illustrates how these advances can be achieved.  The image on the left of this figure shows light intersecting a ruled grating nearly parallel to the groove direction creating an arc of diffraction at the focal plane with dispersion dictated by the displayed grating equation.  On the right, the optical axis is out of the page and the grating grooves are shown projected from the position of the gratings all the way down to the focal plane.  This distance is called the throw and is typically several ($\sim$7--20) meters.  Obtaining high reflectivities of the X-rays necessitates grazing incidence, and in turn, arrays of aligned, stacked gratings.  Such an array can obtain optimal collecting area in the off-plane mount given that the grazing angle of incoming light onto the grating is equal to the half angle of the arc of diffraction exiting the grating surface.  This allows for closely spaced packing geometries that are not available to traditional in-plane reflection grating arrays such as those on \textit{XMM}.  Furthermore, the groove surfaces can be blazed to a triangular profile that preferentially disperses light to one side of zero order.  This increases the signal-to-noise for these orders and limits the size of the detector array.  The blaze angle is chosen to optimize diffraction efficiency at a certain wavelength, typically in the middle of the first order bandpass, which translates to optimized efficiencies at higher orders for shorter wavelengths.  The grating array is then placed with the grooves at a slight angle relative to the optical axis resulting in an $\alpha$ for zero order at the focal plane that equals the $\beta$ of the optimized wavelength.  When $\alpha=\beta=blaze angle$ the array is optimized for diffraction efficiency.

The projection of the grooves in Figure \ref{OPGeom} illustrates the radial distribution of grooves necessary to achieve high spectral resolving power.  The convergence of the grooves matches the convergence of the telescope beam, thus maintaining a constant $\alpha$ at the grating and constant $\beta$ per wavelength at the focal plane, which limits groove profile induced aberration.  The grooves converge to a point at the center of the circle defined by the intersection of the cone of diffraction with the focal plane.  This circle also contains both the telescope focus and the zero order focus.  The overlap of spectra from each of the gratings in the array is achieved by fanning the gratings such that all surfaces project to the diameter of this circle with coincident groove hubs.

While the previous geometric considerations make an off-plane spectrometer an excellent candidate for future X-ray grating missions, key steps still need to be taken in the maturation of this technology. One such step is the development of a process for fabricating and replicating the high groove density (5500--7000 grooves/mm), radially ruled, blazed gratings required for a next-generation instrument. This fabrication process must also be reproducable over large formats ($\sim$100 mm x 100 mm) in order to achieve large collecting areas while maintaining the feasibility of grating alignment and mass restrictions.  These four grating characteristics, radial ruling, blazed profile, high groove density, and large grating formats, represent the major technical challenges to be overcome in next-generation grating fabrication.  In this paper we discuss recent advances in a novel fabrication method that is capable of overcoming these challenges in the coming years.  Specifically, we discuss the fabrication and testing of a ``pre-master'' off-plane grating that exhibits many of the aforementioned characteristics.  We present empirical results from diffraction efficiency and spectral resolving power performance tests.  Finally, we discuss the path forward to increasing the fidelity of these promising gratings.

\section{Off-plane grating fabrication}
\label{sec:1}
Holographic recording has significant heritage with previous off-plane reflection grating studies \cite{McEntaffer2004,Osterman,McECash,Oakley}.  These gratings have been employed in multiple suborbital rocket flights and were explored during the \textit{IXO} development effort. This method of fabrication involves the interference of two recording lasers onto a photoresist coated substrate.  The light sources are placed within the plane containing the grating normal and perpendicular to the eventual groove direction. This method has created parallel grooves at moderate to high densities (4500--5670 gr/mm). Subsequent etching techniques can be used to blaze the grooves while out-of-plane displacements of the recording sources can produce pseudo-radial groove profiles.  However, several limitations currently exist with this technology: higher groove densities require smaller wavelength lasers; reliable recording over large scales has not been accomplished and requires new tooling; and the radial profile is an approximation that results in a non-zero grating induced aberration.  These issues could likely be resolved with significant upgrades to production facilities.  However, uncertainty in the future of the technology development of off-plane grating technology can be removed by identifying suitable fabrication alternatives that take advantage of current, common techniques. 

We have begun to explore a novel grating fabrication method that integrates three common microfabrication techniques: e-beam writing of a photomask, deep ultraviolet (DUV) projection lithography, and anisotropic etching.  Gratings produced with this integrated method can possess a high groove density with a blazed, radial profile and can be replicated over large formats, hence meeting the requirements for a new generation of off-plane reflection gratings.  The first two steps in this process produce a ``pre-master'' exhibiting a high-density, laminar, radial groove profile.  Fabrication of the pre-master using e-beam writing and DUV projection lithography is a technique developed by LightSmyth Technologies.  This method employs computer aided design (CAD) to define the width and location of each grating groove resulting in a series of polygons.  The design is done at four times the scale of the actual grating layout.  The CAD design programs an e-beam writing tool to create a photomask with a very fine mesh spacing of 1 nm.  The mask is prepared from an ultra-flat fused silica plate coated with a film of chromium overcoated with a layer of photoresist.  The resist is exposed in the e-beam writer in accordance with the CAD file pattern, then developed to remove the exposed portions. The groove pattern is transferred into the chromium film using a reactive ion etch (RIE).  The resulting mask exhibits strips of chromium corresponding to the locations of the grooves and transparent areas in the location of the grating trenches.

The mask pattern is subsequently reduced onto a resist coated, single crystal Si wafer via 4x reduction projection lithography resulting in a final spatial resolution of 0.25 nm.  After the resist has been developed the exposed resist is chemically removed.  Next, a reactive ion etch (RIE) is used to transfer the pattern into the Si wafer before the remaining resist is removed by a chemical wash.  Finally, gold film was deposited on the silicon grating using an e-beam coating chamber in a top-down deposition geometry.  The chamber was specially designed by LightSmyth to ensure high fidelity of the gold grating that accurately reproduces the surface corrugation on the Si wafer after $\sim$80 nm of gold deposition. 

\begin{figure}
  \includegraphics[width=5.0in,height=2.93in]{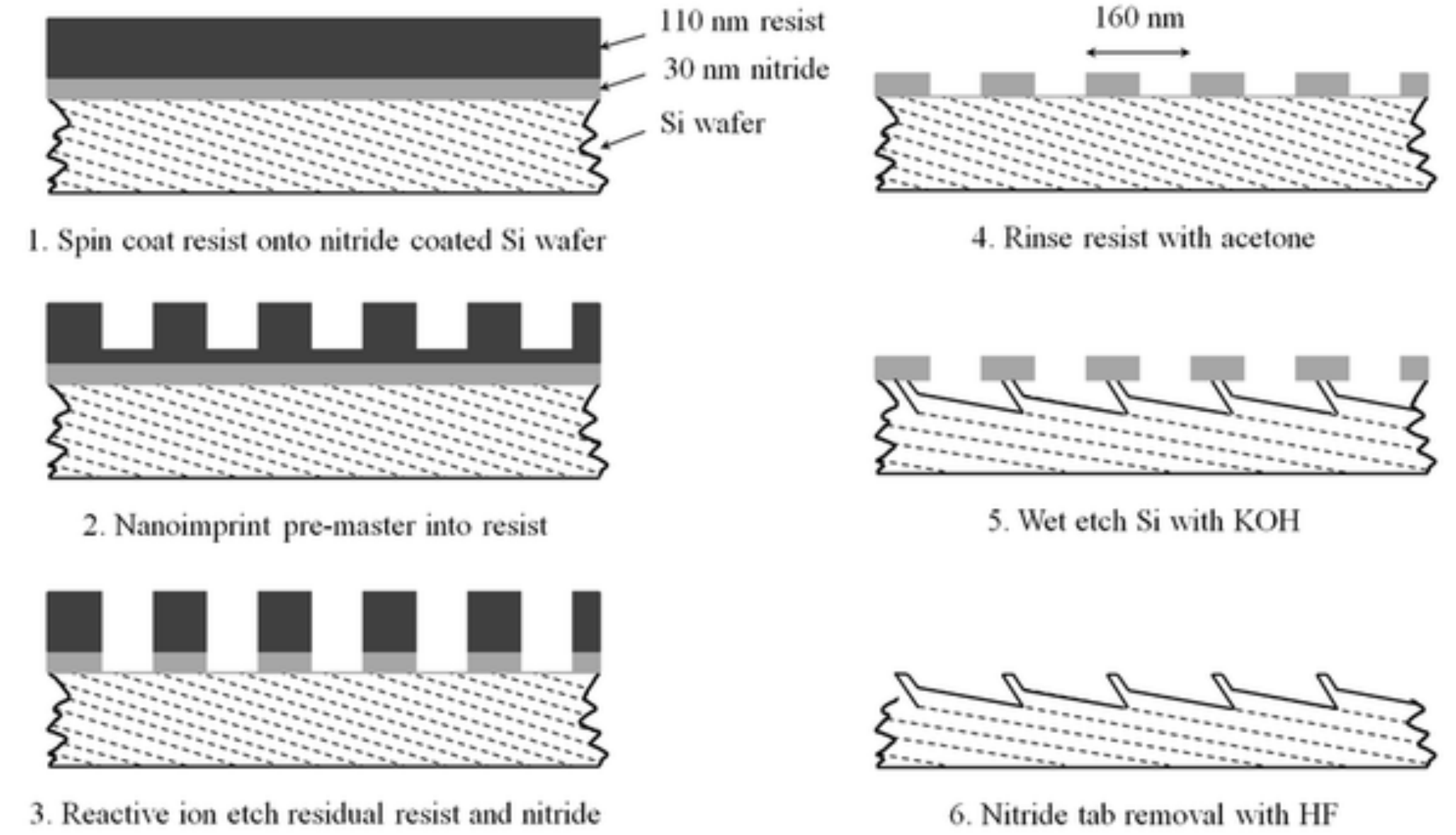}
\caption{Fabrication methodology for blazed, off-plane reflection gratings on Si wafers.}
\label{etch}
\end{figure}

This process yields a very high density grating (up to 7200 gr/mm) with a higher fidelity radial approximation to the grooves in comparison to previous holographic recordings.  The limitation is that the final product is not blazed but a laminar groove profile.  However, common, straightforward, post-processing techniques can sculpt the grooves to achieve the desired blazed profile.  Anisotropic etching is capable of producing gratings with atomically smooth, blazed groove facets but has currently not been used to make a radially fanned groove distribution.  A process to manufacture blazed gratings of moderate groove density (5000 gr/mm) using anisotropic etching was developed at MIT \cite{Chang}, and exploits a preferred direction for KOH etching in single crystal Si. We are currently exploring a similar process which is summarized in Figure \ref{etch}. The novel aspect of this procedure is using nanoimprint lithography to transfer the pre-master pattern into the nanoimprint resist.  An RIE run transfers the pattern into a nitride coating.  Residual resist can be rinsed off with acetone leaving the groove pattern in nitride strips on the Si wafer.  The wafer has been cut off-axis so that the (111) plane is at an angle relative to the plane of the wafer.  This angle represents the blaze angle of the grooves.  A KOH etch preferentially etches along the (111) crystallographic plane leaving atomically smooth, and hence scatter limited, angled facets between the nitride tabs, which are subsequently removed with HF.

By combining laser recording with an anisotropic etching process, it would be possible to exploit the high fidelity radial profile produced by LightSmyth while obtaining a high quality blaze. We have begun the development of such a novel fabrication technique, and present our process here. First, a high density, radially ruled, laminar grating is produced by LightSmyth. This grating serves as a premaster for the manufacturing process. Next, the premaster pattern is imprinted into a layer of resist atop a nitride coated Si wafer. The nitride coated Si wafer will be a single crystal wafer, cut off-axis relative to the [111] plane at the desired blaze angle. The premaster pattern transfer will be accomplished via nanoimprint lithography, a means of embossing on micro/nano-scales, and will result in radially patterned resist. Similar to the anisotropic etching process previously described, an RIE etch is then employed to remove the nitride layer down to the Si wafer, and an RCA clean removes any remaining resist. A subsequent KOH etch acts between the remaining radially ruled nitride tabs, etching down to the (111) plane and producing the desired blaze. The remaining nitride is then removed, and the grating surface is coated with Au, resulting in a blazed, radially ruled, high density grating with excellent X-ray reflectivity at grazing incidence. 

\section{Pre-master characterization}
\label{sec:2}
\begin{figure}
  \includegraphics[width=5.0in,height=1.91in]{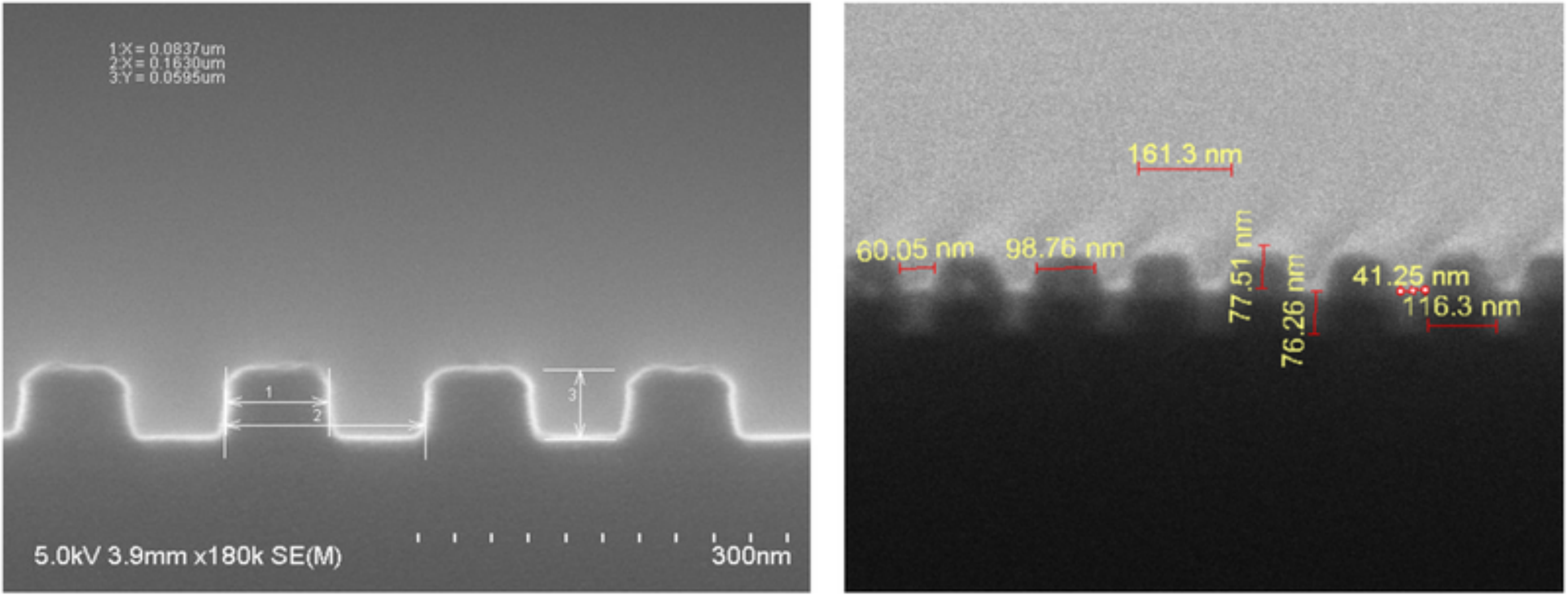}
\caption{\textit{Left} - Cross-section SEM image of the laminar groove profile on single crystal Si.  \textit{Right} - The same Si profile is seen in black with a gray e-beam deposited Au layer on top.}
\label{gratSEM}
\end{figure}

The first step in investigating this integrated process is the fabrication and characterization of the high quality pre-master. To that end, LightSmyth has produced a series of high groove density, laminar gratings over a $25\times32$ mm format. The groove distribution for these gratings is radially ruled to match the convergence of an 8.4 m telescope beam - the focal length of the test optics used during spectral resolution performance testing.  Figure \ref{gratSEM} shows scanning electron microscope (SEM) images of these gratings.  The cross-section image on the left displays an exceptionally clean groove profile with a feature size of $\sim$84 nm, period of $\sim$163 nm, and etch depth of $\sim$60 nm in single crystal Si.  The excellent profile is maintained even after the reflective Au coating has been applied as shown in the image on the right of this figure.  This image shows the same Si profile from the left image as the black substrate on the bottom half of the picture.  The Au coating appears gray and the profile is slightly rounded off at the corners of the groove features causing them to appear more trapezoidal than rectangular.

The input design for the DUV mask uses a perfect radial profile with each groove angularly offset with respect to the next groove by $\sim$4 milliarcsec.  While writing the mask the tool snaps to the closest grid in the 1 nm mesh defining the tool resolution.  The DUV projection step effectively reduces the step size to 0.25 nm.  This is the closest approximation to the radial profile currently achievable.  The result can be effectively thought of as 6 sections along the 32 mm direction ranging from 6024--6042 grooves/mm (165.5--166 nm periods).  These densities are very high and easily achieve the dispersion required for a next-generation off-plane grating.  In addition, the recorded radial profile is the best to date thus achieving two of our goals.

\subsection{Diffraction Efficiency}
Establishing that the pre-master has clean, non-distorted groove profiles and low surface roughness is essential, as any aberrations will be transferred during the subsequent replication process. While the SEM images are strong evidence that the pre-master meets these criteria, performance testing of the pre-master is necessary to provide crucial insight into the processing steps of these gratings. Testing will provide assurance that the initial steps of the process are at a high enough fidelity to produce the quality of grating expected further down the fabrication chain.  As such, a testing campaign to measure the diffraction efficiency of the grating was conducted at the Physikalisch-Technische Bundesanstalt (PTB) soft X-ray beamline located at the BESSY II synchrotron \cite{Laubis,Tummler}. This facility utilizes a grating monochromator sampling the electron storage ring to create a 98\% polarized beam with tunable X-ray energies from $\sim$0.1-1.9 keV.  The test chamber houses staging allowing for sample alignment with six degrees of freedom.  Most importantly for the off-plane configuration, the detector photodiode can scan in two dimensions to sample the entire focal plane. 

\begin{figure}
  \includegraphics[width=5.0in,height=3.08in]{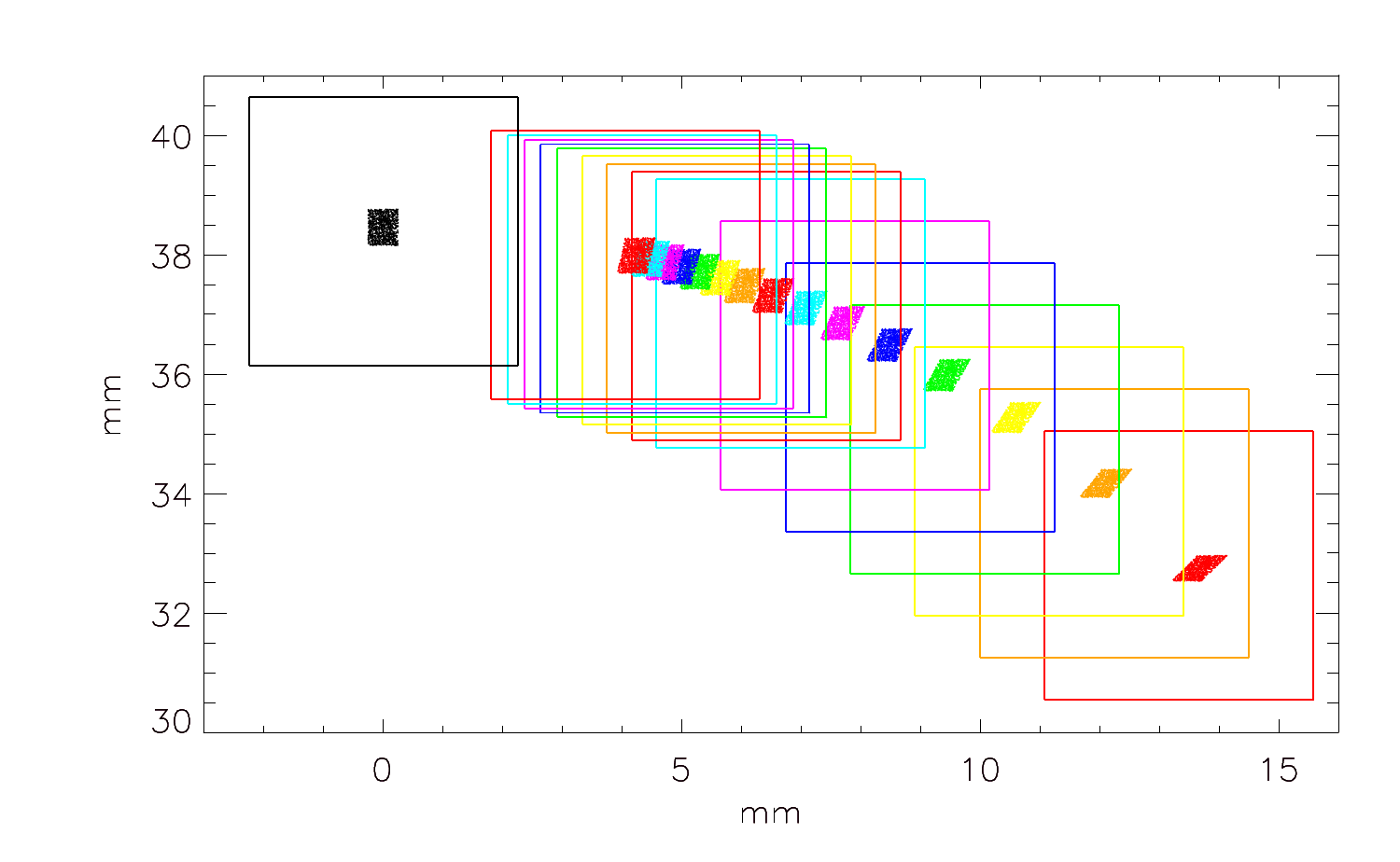}
\caption{The raytrace of the expected arc of diffraction is shown as a series of 16 small scatter plot rectangles with zero order in black and the 15 tested energies from 0.3--1.0 keV in a rainbow of colors.  Associated diode positions for each energy are shown as larger squares with the corresponding color.}
\label{rainbow}
\end{figure}

The grating was tested from 0.3--1.5 keV with 50 eV steps for three graze angles.  Figure \ref{rainbow} demonstrates the testing methodology using a raytrace of the arc of diffraction expected from the 16 energies (including zero order) tested at an incidence angle of 88$^\circ$ (graze angle of 2$^\circ$).  We show the position of the diffracted spots as derived via raytrace, with each spot measuring 0.5 mm horizontally and 0.6 mm vertically (the grating plane is horizontal), which is the typical beam aperture at the position of the test sample.  We overlay squares representing the larger diode area, $4.5\times4.5$ mm.  The positions used for these diode positions are derived from those actually used during testing to show the congruence between the raytrace expectations and the empirical test setup.  The raytraced spots display an aberration that skews the aperture rectangle into a parallelogram.  This effect is caused by the finite size of the beam on the grating leading to path length differences at the focal plane when considering one horizontal limit of the beam versus the other.  It is more noticeable at long wavelengths where dispersion is highest leading to the largest path length differences.  However, given that the size of the diode is much larger than the aperture, the aberration does not effect the efficiency measurements.

\begin{figure}
  \includegraphics[width=6.0in,height=8.0in]{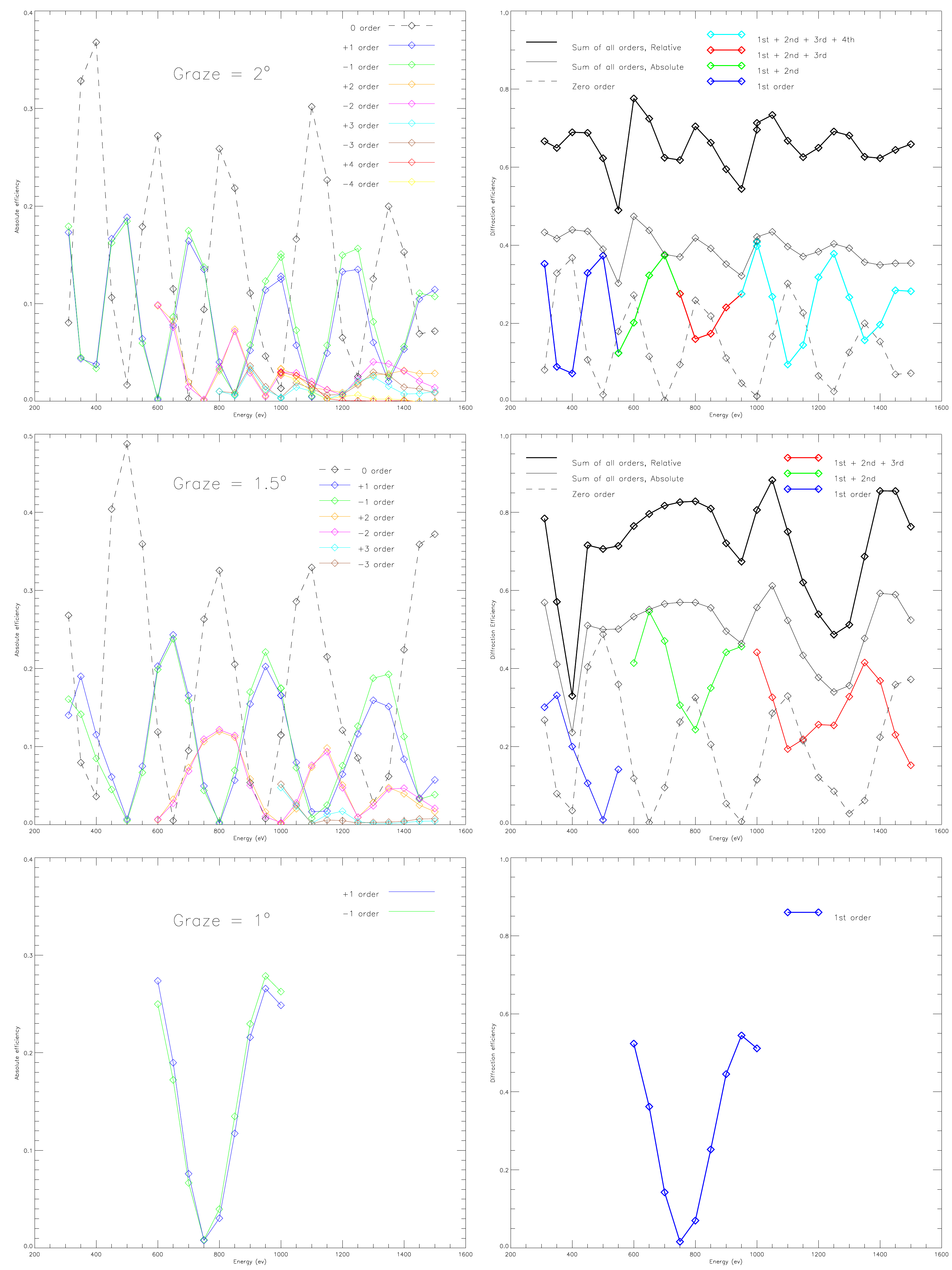}
\caption{Measured diffraction efficiencies at graze angles of 2$^\circ$ (top row), 1.5$^\circ$ (middle row), and 1$^\circ$ (bottom row).  Measurements of individual orders are shown in the left column as absolute efficiencies which are convolved with the reflectivity of Au.  Summed order efficiencies are shown in the right column.  The thin, solid black line shows the total absolute efficiency of all orders, including zero order, while the thick, solid black line shows the total relative efficiency which factors out the effect of the Au reflectivity.}
\label{BessyResults}
\end{figure}

Figure \ref{BessyResults} displays the measured diffraction efficiencies for the BESSY campaign.  Results are shown for graze angles of 2$^\circ$ (top row), 1.5$^\circ$ (middle row), and 1$^\circ$ (bottom row).  The left side of the figure gives absolute efficiencies (diffracted light/incident light with Au reflectivity folded in) at each measured order while the right side shows overall diffraction efficiencies through summation of orders.  It is important to note that the grating was tested at $\alpha = 0$ (light parallel to the grooves).  This led to a limitation on available orders at low energy, which results in only one or two measurable orders over a significant range of our bandpass for this configuration.  The effect of the laminar profile is evident - there are large contributions to zero order, the positive/negative orders contain a nearly equal number of photons, and the diffraction pattern is quite regular and stable over a large range of energies. This symmetry of positive/negative orders is important, as it indicates minimal asymmetry in the groove profile across the grating which will result in a cleaner groove profile being transferred into the resist during the subsequent nanoimprint process.

At 2$^\circ$, the sum of diffracted orders range from 10--40\% in absolute efficiency with 30--40\% being typical.  When considering zero order contributions the total amount of light in the arc of diffraction averages around 70\% after deconvolving the Au reflectivity.  This suggests that up to 30\% of incident light is lost.  This could be due to absorption in the Au coating, or scattering from the surface micro-roughness.  The latter will be discussed below, but some insight to the former can be gained through inspection of the 1.5$^\circ$ results.  At this lower graze angle the relative efficiency averages $>10$\% more total light into the arc of diffraction than that observed at 2$^\circ$ after removing the dependence of the Au reflection efficiency.  At 1.5$^\circ$ the absolute efficiency of diffracted orders is consistently above 40\% when multiple orders are present and greater than 50\% at 650 eV.  At a 1$^\circ$ graze angle, of the few energies we were able to measure we again find efficiencies that are well above 50\% at 600 eV and from 950--1050 eV.  Given the small graze angle and the $\alpha=0$ configuration, only $\pm1$st order are available (zero order was present but not measured).  Overall, the gratings consistently produce efficiencies above 30\% and up to 55\%.

Due to a lack of time at the facility, we were not able to test at alpha angles other than zero.  At higher alphas we would be able to probe higher order diffraction at lower energies.  However, large alphas are not optimal given the existing laminar profile which leads to significant groove shadowing and large diffraction angles off of steep side walls.  This underscores the necessity of the blazing process.  Blazed grooves will allow for maximal illumination of the grooves at a constant angle (alpha = blaze angle) resulting in the ability to test at high alpha.  At large alphas the zero order image is closer to the plane of the grating (see Figure \ref{OPGeom}).  Therefore, not only is dispersion preferentially on one side of zero order, it disperses to larger betas and higher orders as well.  Ultimately this results in minimal zero order contribution, and additional contributions from higher orders.  In such a configuration the sum of diffracted orders resembles the solid black lines on the right side of Figure \ref{BessyResults}. The effect of a dominant diffraction contribution is already seen over small energy ranges: around 1000 and 1250 eV at 2$^\circ$; around 850, 950, and 1350 eV at 1.5$^\circ$.  At these higher energies there are typically multiple diffraction orders with minimal contributions from zero order, thus approximating what we would see more ubiquitously given blazed facets on these gratings.

The 40\% absolute efficiency limit is an important goal.  Previous studies pointed toward the possibility of achieving this level for the off-plane configuration \cite{McEntaffer2004}. Using these initial tests, conceptual designs for instruments onboard future NASA missions such as the \textit{Warm-Hot Intergalactic Medium Explorer} (\textit{WHIMEx}) and \textit{IXO} XGS \cite{Bautz,McEntaffer2011} have used 40\% diffraction efficiency as the basis for determining spectrometer properties.  The assumption of obtaining this level flows down to the size of the array, the number of gratings, their size/mass/material, etc.  In short, the entire design of the grating assembly is determined by their efficiency.  These comprehensive BESSY tests show that the efficiencies assumed in previous IXO designs are obtainable, even with an unoptimized pre-master.

\begin{figure}
\includegraphics[width=3.85in,height=2.05in]{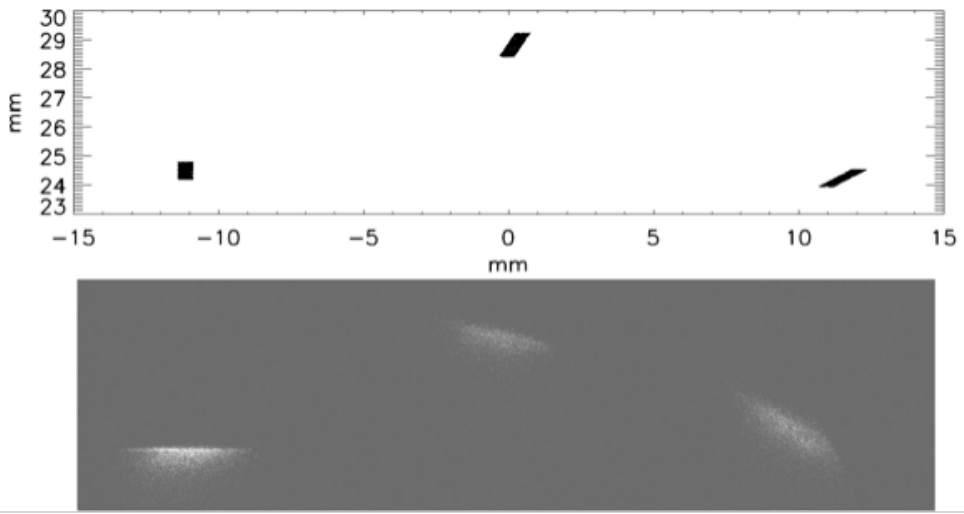}
\caption{The upper plot is a raytrace of the pre-master in the BESSY beamline onto the CCD focal plane.  The spot on the left is the zero order image of the slit using 1.9 keV with 1st and 2nd order diffracted toward the right.  The lower plot is a CCD image of the actual arc of diffraction for 1.9 keV X-rays and shows excellent agreement with the raytrace predictions.  The physical horizontal size of the CCD is 26 mm.}
\label{ccd}
\end{figure}

While at BESSY, in addition to efficiency testing, the diffracted beam was also directed onto a CCD camera to image the arc of diffraction.  Given the small number of access ports on the PTB test chamber, the camera was placed along the beam axis thus limiting the graze angle to 0.25$^\circ$ and our spectral range to only the highest energies.  Even so, we were able to run the monochromator at 1.9 keV and image two diffraction orders along with zero order on a single CCD.  As shown in Figure \ref{ccd}, the diffraction properties of the grating match the theoretical raytrace closely, providing further verification on the high quality of the groove profile.  At this low graze angle the aperture no longer appears rectangular, but more diffuse.  Close inspection of these data show no obvious sign of scatter off of the grating surface which should be preferentially in the in-plane or vertical direction \cite{McEntaffer2004}.  Therefore, we see no sign of scatter leading to a loss of incident light into the arc of diffraction as suggested in the efficiency tests.  We have also measured scatter off of these pre-master gratings at the University of Iowa X-ray Test Facility.  Up to graze angles of 2.5$^\circ$, we see no measureable amount of scatter. 

\subsection{Spectral Resolving Power}
In addition to diffraction efficiency, the science goals of future observatories require high spectral resolving power for the next generation of gratings.  For the \textit{IXO} project this translated to $\lambda/\Delta\lambda>3000$ over a 0.3--1.0 keV band.  This energy range is limited on the low end by the response of the CCD camera.  The high end will be determined by comparing the resolutions and sensitivities between gratings and microcalorimeters, but will most likely lie between 1.0--2.0 keV.  The major hurdle to obtaining high spectral resolution in the off-plane mount is producing a high fidelity radial groove profile.  As explained in section \ref{sec:1}, the pre-masters approximate this profile by using several parallel sections with groove density increasing in the direction toward the focal plane.  The pre-master was tested for spectral resolving power at the Marshall Space Flight Center (MSFC) Stray Light Facility (SLF).  This is a 100 m beamline utilizing a Manson electron impact X-ray source at one end with a large 3 m diameter, 10 m long cylinder at the opposite end to house the experiment.  We utilized a Wolter-I telescope fabricated at Goddard Space Flight Center (GSFC) to create the focused beam \cite{Zhang}.  Glass slumping onto precision mandrels is used to create the parabola/hyperbola pairs.  We used a Technology Development Module (TDM) that has three pairs of these mirrors aligned to create a $<$10 arcsecond half power diameter (HPD) focus at a nominal focal length of 8.4 m.  These mirrors only cover $\sim$30$^\circ$ of azimuth thus resulting in a subapertured beam with a much tighter focus in one dimension.  We subapertured the beam even further given that our gratings are only 25 mm wide ($\sim$6$^\circ$ of azimuth).  The mirror focal length at the finite conjugate (90 m source distance) is $\sim$9.3 m.

\begin{figure}
  \includegraphics[width=5.0in,height=3.33in]{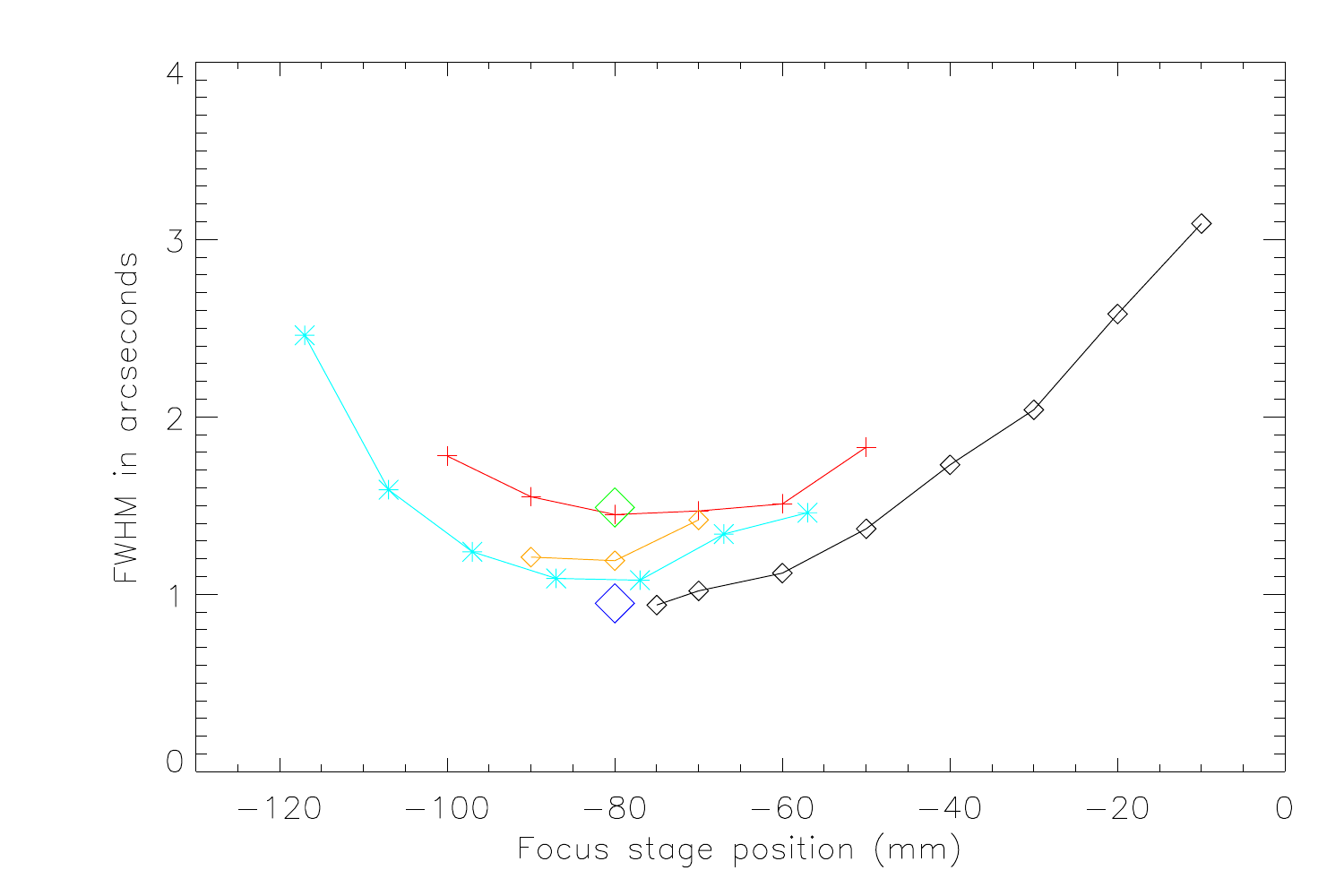}
\caption{Zero order focus results:  Black diamonds - focus run on Day 1 with Manson electron beam current at 0.16 mA; Red crosses - focus run on Day 2 with Manson beam current at 0.5 mA; Cyan asterisks - focus run on Day 3 with Manson beam current at 0.16 mA; Green and Blue diamonds - focus checks on Day 4 with Manson beam at 0.5 mA and 0.16 mA, respectively; Orange diamonds - focus check on Day 5 with Manson beam at 0.16 mA.}
\label{ZeroFoc}
\end{figure}

The test began with finding a focus for zero order.  Focus runs were performed each day and the results are summarized in Figure \ref{ZeroFoc}.  The best focus obtained in the dispersion direction was 0.94 arcseconds full width at half maximum (FWHM) on Day 1.  Even though the quality of the focus varied slightly day-to-day, a much larger effect was observed by changing the electron beam flux at the Manson source.  The impact source utilizes a high potential field (several kV) to accelerate electrons into a solid anode, in this case Mg, to create fluorescence lines.  Such resolution tests are often flux limited so it is typically a necessity to maximize the source flux.  This is done through increasing the electron beam current of the Manson, or in other words the number of electrons allowed to impact the anode.  As the stream of electrons increases, so does the repulsive Coulomb force in the beam causing the size of the impact spot to increase as well.  Therefore, at high flux one typically observes a larger impact spot on the anode.  We found that the GSFC TDM is capable of resolving the impact spot at 90 m (1 arcsecond $\sim$0.44 mm) even at the lowest beam current setting (0.16 mA), resulting in a diffuse source.  This applies a systematic limit to the spectral resolution that is necessarily factored into the results.

\begin{figure}
  \includegraphics[width=5.0in,height=2.0in]{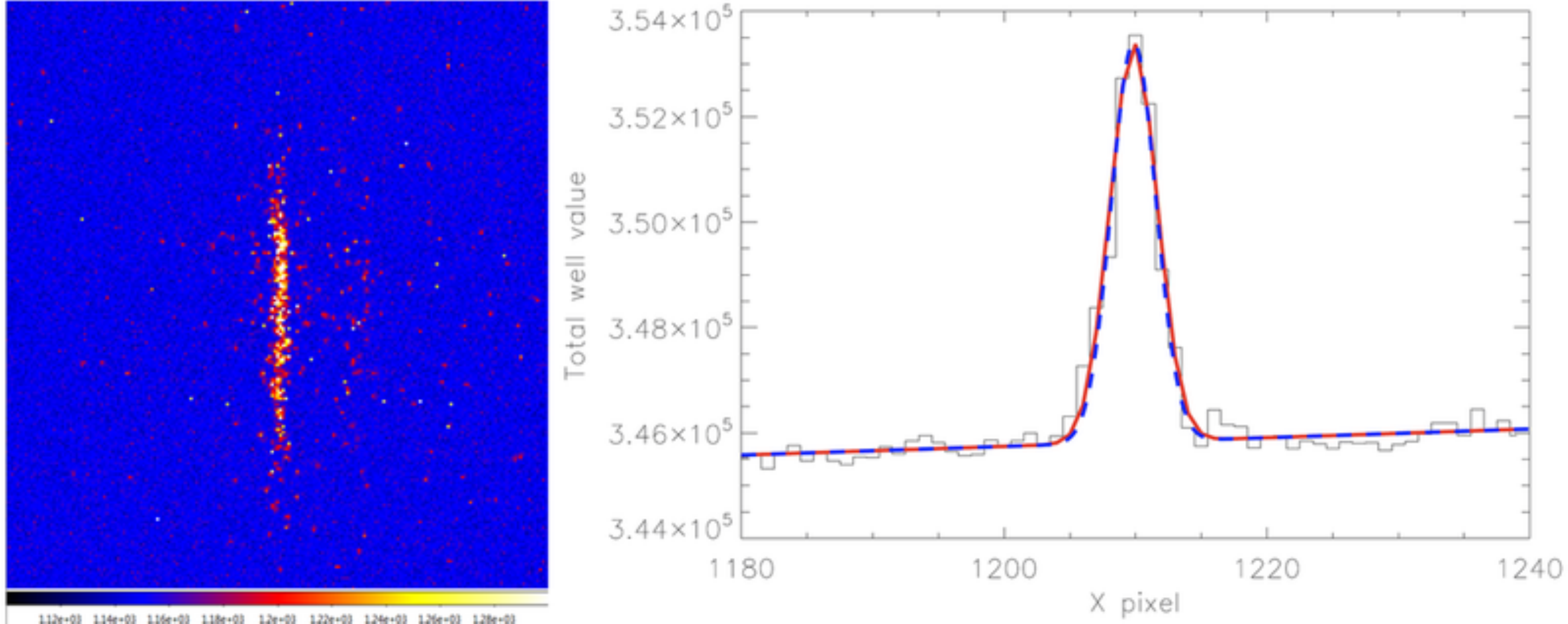}
\caption{\textit{Left}--CCD image ofthe first order Mg K$\alpha$ line.  \textit{Right}--Histogram of all CCD counts with best fit overlaid in red and theoretical expectation in dashed blue.}
\label{FirstOrder}
\end{figure}

On Day 2 we performed repeated measurements of the first order lines.  In general, these measurements were made at high beam current given that we were concentrating on imaging and alignment.  Post alignment, we turned down the beam current and obtained our best first order resolution.  An image and histogram of the line are given in Figure \ref{FirstOrder}.  The histogram is fit with a Gaussian profile (red curve) to obtain a FWHM of 1.26 arcseconds which translates to 4.2 pixels with each pixel measuring 13.5 $\mu$m.  To determine the grating resolution we must first calculate the dispersion.  Given the average groove density of 6033 grooves/mm and a mirror-to-grating distance of 785 mm, we get a dispersion of \AA/mm$=10^7$/(groove density$\times$throw)$=0.19$.  Therefore, the line width is 0.011 \AA  giving a resolution of 897 ($\lambda/\Delta\lambda$) at 9.89 \AA (Mg K$\alpha$).  This also matches the x/$\Delta$x (dispersion distance/physical line width) relation given that the line is positioned 51.0 mm away from zero order with a 4.2 pixel (0.0567 mm) width giving a resolution of 899.

Also shown in Figure \ref{FirstOrder} is the theoretical expectation of the first order line, represented by the dashed blue line.  The theoretical resolution must factor in both the line spread function of zero order and the expected aberration from the radial groove approximation.  The aberration caused by this approximation has been raytraced using Interactive Ray Trace (IRT; v2.7).  The raytrace calculates a grating contribution of 0.47 arcseconds (FWHM) to the spectral line spread function.  Therefore, given a point source and a scatter dominated telescope focus, this grating is capable of achieving theoretical first order resolutions above 2500.  However, the source is resolvable and the telescope has figure errors that contribute to the telescope HPD such that the zero order focus at the time of first order testing was 1.1 arcseconds.  Summing these errors results in a 1.2 arcsecond FWHM which as seen in Figure \ref{FirstOrder} characterizes what we measured quite well.  Even so, theory suggests a slightly tighter FWHM than that measured.  This small difference does yield a slightly higher theoretical resolution of 996, but practically is only 1/5 of a pixel.  There may be several factors that are contributing to this small difference including: an assumption that the lines are perfectly vertical on the CCD, split events contributing to the width, fluctuating Manson flux levels, unstable mirror focus with temperature, and/or a small deviation in the optical focal plane.  Given these factors the measured resolution is remarkably close to the theoretical limit.

\begin{figure}
  \includegraphics[width=4.0in,height=3.59in]{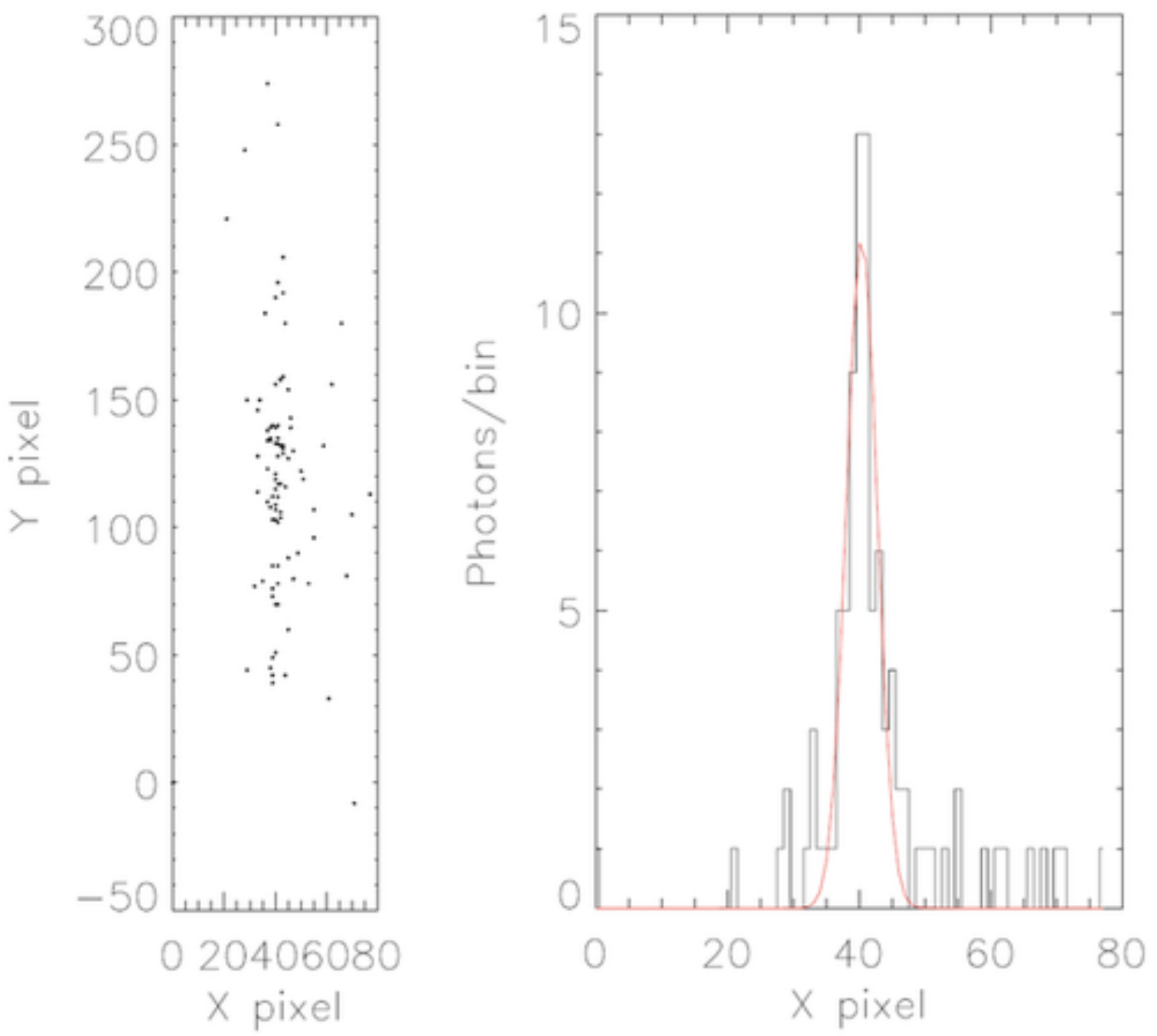}
\caption{\textit{Left}--Second order line reconstructed from only $>5\sigma$ events.  \textit{Right}--Histogram of these photons with best fit Gaussian overlaid in red.}
\label{SecondOrder}
\end{figure}

Second order testing occurred on Day 5 when the best zero order focus obtained was 1.19 arcseconds at low beam current.  Sampling of the focal plane was limited by the maximum displacements of the linear stages controlling the CCD position.  Therefore, to image higher order spectral lines we had to apply an alpha to the grating to position the arc of diffraction at a vertical and horizontal position consistent with the linear stage ranges.  This necessitated a larger graze angle as well and ultimately resulted in much lower efficiency in the second order line.  We therefore operated at high Manson flux, 0.4 mA, resulting in an intrinsically larger spot.  Given the low number of counts at the focal plane, we operated in photon counting mode which takes 270 individual 30 second exposures over 3.5 hours.  When analyzing this data cube we only consider counts that are 5$\sigma$ above the median value of an individual 30 second exposure, where $\sigma$ is the standard deviation pixel value in the individual image.  The charge in the surrounding 8 pixels is summed for each pixel above this 5$\sigma$ threshold.  This results in a total photon count of 93 photons.  The resulting line and histogram of counts are shown in Figure \ref{SecondOrder}.  While this number is lower than optimal, it represents only those counts that are well above the background.  The photon counts can be significantly increased with lower CCD backgrounds (typical background pixel $\sim$90 counts; typical photon pixel $\sim$150 counts with splits decreasing significance per pixel) and more optimal grating configurations.  The width of the 2nd order line is 5.48 pixels resulting in a 1.64 arcsecond focus.  Given the grating dispersion this results in a resolution of 1375.  Again, this matches well with the observed x/$\Delta$x dispersion of 98 mm/0.074 mm = 1324.  It is expected that a second order line will achieve a resolution $2\times$ that of 1st order, i.e. $\sim$1800.  Instead, the measured 2nd order line width is 1.64 arcseconds, but this may be expected given our operation at 0.4 mA of beam current.  We did not measure the focus at this source level, but did see 1.2 arcseconds at 0.16 mA earlier in the day and 1.5 arcseconds at 0.5 mA the day before.  Using the latter convolved with the expected grating induced aberration results in a 1.6 arcsecond beam.  Therefore, we once again found the resolution to be limited by the source size.  Furthermore, the 2nd order data were taken just after the chamber vacuum was interrupted by a cryo pump self-shutdown resulting from a lack of liquid nitrogen.  Significant temperature deltas were observed on this day possibly resulting in poorer telescope resolution on top of the finite source size.

\section{Summary and Future Work}
In the context of future NASA X-ray missions, X-ray grating spectrometers must be able to achieve high throughput and high spectral resolution.  Our goal is to realize this performance by utilizing off-plane reflection gratings.  We have identified a novel grating fabrication technique that is capable of producing the radial, blazed, high density profile that is required to obtain this goal.  The unique first step involves creating a pre-master laminar grating in single crystal Si through e-beam production of a mask with subsequent DUV lithography.  This step has been accomplished and has created a radial profile grating matching an 8.4 m focal length beam with very high groove density, 6033 grooves/mm.  This pattern will be transferred via nanoimprint lithography to a resist coated master that will undergo subsequent etching steps to ultimately create the blazed groove profile.  These latter steps have been performed by others and are well understood leaving just the qualification of the pre-master as an open issue.  We have performed this qualification and have found that the pre-master performance is consistent with our goal throughput and resolution numbers.  The non-blazed, laminar profile consistently diffracts $>30$\% of incident light into usable spectral orders and upwards of 55\% in some cases.  When multiple orders are present the sum of orders is typically $>40$\%.  Blazing this profile should produce a consistent 40\% diffraction efficiency.  Furthermore, we have obtained theoretical spectral resolutions using a high-quality X-ray optic.  We found resolutions of $\sim$900 in first order and $>$1300 in second order.  These values are limited by the finite source size used in the experiment with much higher resolutions theoretically and practically possible.

\begin{figure}
  \includegraphics[width=5.0in,height=1.83in]{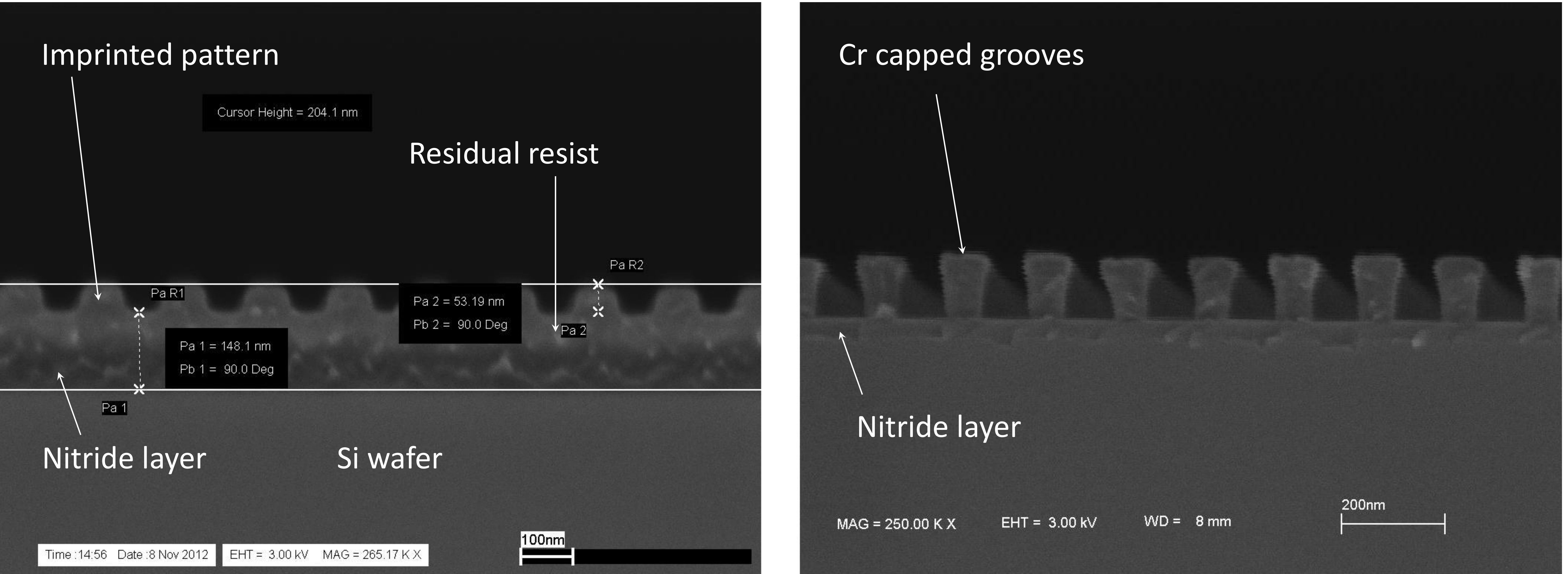}
\caption{\textit{Left}--SEM image of the imprinted pattern in the resist of the secondary wafer.  \textit{Right}--The same wafer after removal of the residual resist via oxygen RIE.}
\label{imprint}
\end{figure}

The next step will be to test the performance of blazed, radial gratings.  The first step in creating this grating, nanoimprint lithography \cite{Chou,Gao}, has been accomplished by our industrial partners at Nanonex using our pre-masters as shown in Figure \ref{imprint}.  The SEM image on the left shows the pre-master pattern imprinted into a thick resist layer on a nitride coated Si wafer.  The residual resist is removed by masking the tips of the grooves with Cr and oxygen plasma etching via RIE.  Etching of the nitride and a wet etch of the Si will create the necessary blazed profile.  Performance testing for throughput and resolution on these samples will verify this process as a leading tool in next generation diffraction grating fabrication.

\begin{acknowledgements}
This work was supported by NASA grants NNX12AF23G and NNX12AI16G.  We would also like to acknowledge internal funding from the University of Iowa in support of Casey DeRoo.  Special thanks are due to several people including Christian Laubis and his terrific support crew at PTB as well as James Carter and Bill Jones at MSFC for support of the resolution tests.
\end{acknowledgements}

\end{document}